\documentclass[10pt,letterpaper]{article}
\usepackage{opex3}
\newcommand{\ket}[1]{\ensuremath{\left|#1\right\rangle}}
\newcommand{\Fig}[1]{Fig.~\ref{#1}}

\begin{document}

\title{Post-selection free, integrated optical
source of  non-degenerate, polarization entangled photon pairs}

\author{Harald Herrmann$^{1,2\ast}$, Xu Yang$^1$, Abu Thomas$^2$, Andreas Poppe$^3$ \\ Wolfgang Sohler$^2$, and Christine
Silberhorn$^1$}

\address{$^1$ Integrated Quantum Optics Group, University of
Paderborn, \\ Warburger Str. 100, 33095 Paderborn, Germany\\
$^2$ Applied Physics / Integrated Optics Group, University of Paderborn\\
Warburger Str. 100, 33095 Paderborn, Germany\\
$^3$ AIT Austrian Institute of Technology GmbH\\
Donau-City-Str. 1, 1220 Vienna, Austria}

\email{$^{\ast}$h.herrmann@physik.uni-paderborn.de} 



\begin{abstract}
We present an integrated source of polarization entangled photon pairs in the telecom regime, which is based on type
II-phasematched parametric down-conversion (PDC) in a Ti-indiffused waveguide in periodically poled lithium niobate. The domain grating -- consisting of an interlaced bi-periodic structure -- is engineered to provide simultaneous phase-matching of two PDC processes, and enables the direct generation of non-degenerate, polarization entangled photon pairs with 
a brightness  of $B=7\times10^3$ pairs/(s$\times$mW$\times$GHz). 
The spatial separation of the photon pairs is accomplished by a fiber-optical multiplexer facilitating a high compactness of the overall source. Visibilities exceeding 95~\% and a violation of the Bell inequality with $S=2.57\pm0.06$
could be demonstrated. 
\end{abstract}



\section{Introduction}

The capability to distribute entanglement to remote locations is a precursor for future quantum cryptography systems and  more advanced quantum networks.
Of particular interest are entanglement-based quantum communication systems compatible with existing standard fiber telecom networks.
In such networks entanglement is distributed between different parties by means of entangled photons favourably in the 1.55~$\mu$m telecom band.
Therefore, the development of telecom compatible, compact and reliable sources of entangled photons is a prerequisite to pave the way for a successful implementation of future quantum 
information technologies into real world applications.  

Different approaches have been demonstrated to generate entangled photon pairs. Among all possible methods parametric down-conversion (PDC) in $\chi^{(2)}$ based nonlinear
materials is frequently preferred due to comparatively high efficiency and relatively simple generation scheme. In particular high efficiency and source brightness can be obtained by taking advantage
of integrated optical structures. Keeping the interacting waves confined within distinct spatial optical modes of a waveguide  provides several order higher efficiencies compared to corresponding bulk optic devices. \cite{Suhara_Review,Gisin,Fiorentino}.   
 
Efficient photon pair generation by PDC in waveguides has been demonstrated by several groups (see e.g.\ \cite{Gisin, Fiorentino,Uren, Fujii, Tanzilli1,Suhara1}). 
The most established material is lithium niobate, because mature waveguide fabrication technologies exist and ferroelectric domain inversion enables  periodic poling for quasi-phase-matching (QPM) making this material an ideal candidate for quantum optic devices~\cite{Tanzilli-Review}. In particular Ti-indiffused waveguides in periodically poled LiNbO$_3$ (PPLN) 
have been applied to generate entangled orthogonally polarized photon pairs
using a type II QPM process \cite{Tanzilli1,Suhara1,Suhara_Sequential, Tanzilli2}. 

The operational principle of most of these devices relies on 
photon pair generation at or at least close to degeneracy, i.e.\ the wavelength of both photons of the pair is identical. A subsequent spatial splitting 
of the pair has been applied using a standard 50:50 beam splitter or a fiber coupler.   This has a severe drawback because only half of the generated pairs are spatially separated. This does not only reduce the effective generation efficiency, but also necessitates a post-selection process for the extraction of the entanglement which prohibits many applications like e.g.\  entanglement distillation.

To overcome this drawback a nondegenerate PDC scheme exploiting a biperiodic poling structure has been proposed~\cite{Thyagarajan}.  The biperiodic poling enables simultaneous type II phase-matching for two PDC-processes. The  poling periods must be chosen that in the first process a pair with a TE-polarized photon at $\lambda_1$ and a TM-polarized photon at $\lambda_2$ and in the second process a pair at the same wavelengths but opposite polarization is generated.  

The practical implementation of such a scheme  has  been demonstrated in previous work. In \cite{Suhara_Sequential} a PPLN wave\-guide is used with one fixed poling period. To obtain phase-matching for two (slightly) different processes, one part of the sample was covered with a dielectric overlay resulting in a shift of the propagation constants of the guided modes. 
In  \cite{Edamatsu} a bulk optical PPLN crystal with a biperiodic poling structure consisting of a first section with poling period $\Lambda_1=9.25 \mu$m and a second section with $\Lambda_2=9.5 \mu$m  has been used to generate the polarization entangled pairs. Both approaches used a {\em sequential} arrangement, i.e.\ the first PDC process occurs in a first section followed by a further section for the second PDC process. In this way the processes are spatially separated with the disadvantage
that any imbalance can lead to a remaining distinguishabilty and, thus, deteriorates the quality of entanglement.

An advanced approach exploits the whole interaction length for both processes~\cite{Thomas1}\cite{Thomas2}.  This scheme applies {\em interlaced} sections of different poling periods to obtain enhanced indistinguishability.  Any variation of process parameters, e.g.\ pump depletion due to waveguide losses, will simultaneously effect both processes and, thus, preserves the indistinguishability. In this paper we present an integrated source exploiting this scheme. Our source, which provides polarization entangled pairs in the telecom regime,  essentially consists of a PPLN waveguide with a specifically tailored ferroelectric domain pattern for the generation of photon pairs in combination with standard fiber optical components.

The paper is structured as follows: In the next section we  discuss the principle of operation followed by a discussion of the design, fabrication and classical characterization of the integrated optical chip in Sec.\ref{PDC_Section}. 
Details of the entire entangled source and its performance are given in Sec.\ref{entanglement}.

\section{Principle of Operation}
The basic principle of our entangled photon pair source relies on the generation of
photon pairs by PDC in a periodically poled lithium niobate waveguide.
In a PDC process a pump photon decays into a  photon pair, typically labeled as signal and idler photons. We generate orthogonally polarized photon pairs by
applying a type II phase-matched PDC process which exploits the non-linear coefficient $d_{31}$ instead of the frequently used type I
process. This uses the larger $d_{33}$ coefficient, but provides only extraordinarily polarized photons. Besides energy conservation 
(quasi-)phase-matching is required for an efficient process:
\begin{equation}
\beta_p=\beta_s+\beta_i+\frac{2\pi}{\Lambda}
\end{equation}
with $\beta_j ~(j=p,s,i) $ being the wave numbers of the pump, signal and idler wave, respectively. The parameter $\Lambda$ is the period of the ferroelectric domain pattern.

%
\begin{figure}[htbp]
\begin{center}
\includegraphics[width=8cm]{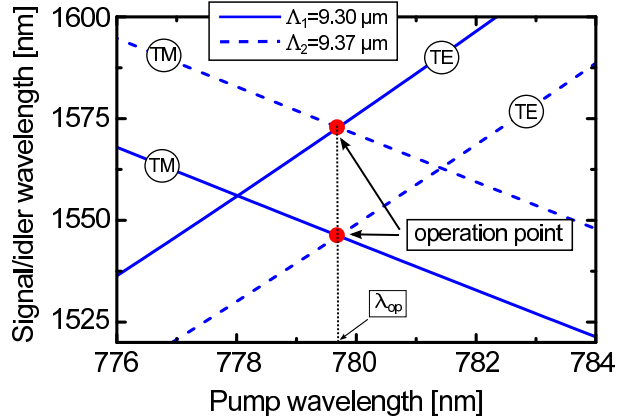}
\caption{\label{PhaseMatching}Calculated phase-matching characteristics of  type II phase-matched PDC processes in a PPLN
 waveguide assuming a poling period of $\Lambda_1=9.30 ~\mu$m (solid lines) and  $\Lambda_2=9.37 ~\mu$m (dashed curves), respectively. The operation point for
the generation of polarization entangled pairs is at the intersection of the phase-matching characteristics of the two periodicities. }
\end{center}
\end{figure}


In \Fig{PhaseMatching} the calculated phase-matching characteristics of our type II process is shown, where
the signal and idler wavelengths are plotted as function of the pump wavelength for two different poling periods $\Lambda_1=9.30~\mu$m (solid lines) 
and  $\Lambda_2=9.37~\mu$m (dashed lines).  For a fixed single poling period the curves have a cross-like shape. Thus, at the intersection of the curve wavelength degenerate photon pairs can be generated. But, as pointed out in the introduction, polarization entangled pairs can only be generated with a beam splitter followed by a post-selection process.

With a biperiodic pattern non-degenerate polarization-entangled pairs can directly be generated: The Ti:PPLN waveguide 
is poled  with two two periodicities $\Lambda_1$ and $\Lambda_2$, such that  the photon pairs can be generated by two different PDC processes, one corresponding to $\Lambda_1$ and the other to $\Lambda_2$. Polarization entangled pairs are obtained if the device is operated at a pump wavelength where the TE emission of $\Lambda_1$
coincides with the TM emission of $\Lambda_2$ and vice versa (see \Fig{PhaseMatching}). The generated state is then given by:

\begin{equation}
\ket{\psi}=\frac{1}{\sqrt{2}}\left[\ket{H}_{\lambda_s}\ket{V}_{\lambda_i}+e^{i\Phi}\ket{V}_{\lambda_s}\ket{H}_{\lambda_i}\right]
\end{equation}
with $\lambda_s$ and $\lambda_i$ being the wavelength of the signal and idler photons, respectively. The states $\ket {H}$ and $\ket{V}$ represent TE and TM polarized single photon states, respectively,  and $\Phi$ is the relative phase between the photon pairs emerging from the waveguide. 

This generation scheme can be implemented using a sequential arrangement of sections with two different poling periods, i.e.\ a first section of  poling period $\Lambda_1$ is followed by a second section with  $\Lambda_2$.  These two sections must be perfectly balanced exhibiting the same PDC properties (efficiency, bandwidth, ...) for high quality entanglement. Practically, however, an imbalance can be hardly avoided. For instance, waveguide losses cause a depletion of the pump power along the waveguide resulting in a weaker pumping in the second section. To overcome this problem a homogeneous distribution of the 
two PDC processes along the entire waveguide structure is required. 
Using the overall waveguide length for the two processes provides also a further advantage: The spectral bandwidth of the generated photons scales reciprocal with length, thus, the predicted PDC bandwidth is significantly smaller because the entire poled area is used for both processes.   

To approximate such a homogenous distribution over the whole waveguide length, our implementation uses an {\em interlaced} scheme which consists of a sequence of short sections with alternating poling periods. Details of the practical implementation of this scheme will be given in the following section.

\section{\label{PDC_Section}PDC waveguide source}
\subsection{Waveguide design and fabrication}
\begin{figure}[htbp]
\centering\includegraphics[width=8cm]{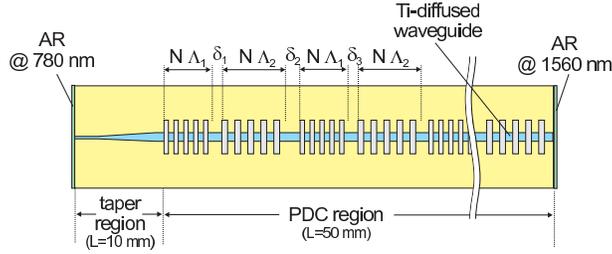}
 \caption{\label{Sample}Detailed design of the integrated PDC source with interlaced poling pattern.}
\end{figure}

The detailed structure of the integrated optical PDC source is shown schematically in \Fig{Sample}. The 60 mm long waveguide was fabricated by an indiffusion of a   90~nm thick Ti-stripe into the 0.5~mm thick Z-cut LiNbO$_3$ substrate at 1060 $^o$C for 9 h. The stripe width in the interaction region is 7 $\mu$m. This fabrication process provides single mode waveguides in the telecom range for both polarizations with typical waveguide losses of about 0.1 dB/cm.  For the pump wave at $\lambda\approx 780~$nm the waveguide is multi-mode. To improve the coupling into the fundamental mode, a taper region is used in which the stripe width increases linearly from about $3~\mu$m to $7~\mu$µm over 10~mm length.

Subsequent to waveguide fabrication the periodic poling was done by field assisted domain inversion. The poling pattern is engineered to provide the interlaced structure: A first section  with $N=10$ domain periods with periodicity $\Lambda_1=9.30~\mu$m is followed by a similar section with periodicity $\Lambda_2=9.37~\mu$m.  This sequence is repeated  over the whole 50~mm interaction length. The spacings $\delta_i$ between the different sections are chosen in a way that the overall spacing between sections with the same periodicity is a respective integer multiple of this periodicity.

%

After polishing the end-faces AR coatings were deposited. They consist of single quarter-wave layers of SiO$_2$ to mimize reflection loss of the pump at the input and of the generated PDC photons at the output. The transmission (measured using a witness sample) is $T\approx95\%$ for the pump at the input face and $T>96~\%$ for signal and idler at output face.

\subsection{Classical characterization}

\begin{figure}[htbp]
\centering\includegraphics[width=7cm]{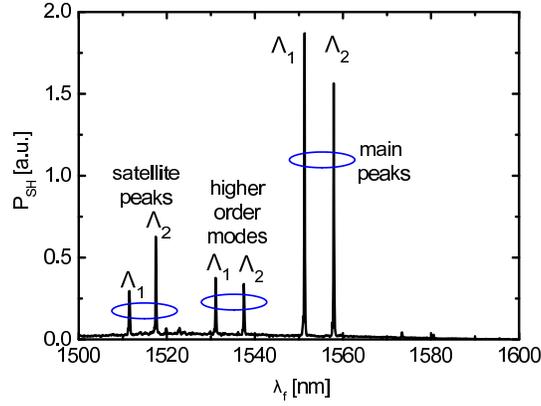}
 \caption{\label{SHG_Spectrum}Measured second harmonic power versus wavelength of the fundamental wave. Details are described in the text. }
\end{figure}

As a first step we set up an experiment to test and characterize the integrated PPLN sample by means of second harmonic (SH) generation. For this purpose light from an extended cavity laser (ECL) tunable  in the telecom regime was coupled into the waveguide. A linear input polarization at 45$^o$ was chosen to excite TE and TM waveguide modes simultaneously.  \Fig{SHG_Spectrum} shows the measured SH power as function of the ECL wavelength. Within the scanned wavelength range three pairs of peaks can be recognized. The dominant pair is the wanted one due to SH generation of the fundamental modes in the sections with $\Lambda_1$ and $\Lambda_2$ periodicities, respectively. The pair in the middle of the spectrum is caused by SH generation resulting from a coupling between the fundamental wave\-guide modes (TE$_{00}$ and TM$_{00}$) to the higher mode TE$_{20}$ at the SH wavelength.  The additional pair at the shortest wavelength - and a similar pair on the long wavelength side outside of the measurement window - are satellite peaks arising from the segmentation of the poling pattern. These satellites occur because the regular interlacing of segments with different poling periodicities results in a QPM superlattice providing additional wave vectors for phase-matching.

%

From our SH measurements we determine a conversion efficiency $\eta_{SH}=P_{SH}/(P_{TE}\times P_{TM})\approx 0.7~\%/W $, where $P_{SH},P_{TE}, P_{TM}$ denote the power of the generated SH wave and the power of the fundamental mode in TE- and TM-polarization, respectively. Using this SH efficiency the efficiency of the PDC process, defined by the ratio of the generated PDC photon pairs to the coupled number of pump photons, can be estimated to be $\eta_{PDC}\approx 1.4\times 10^{-10}$ according to the model described in \cite{Kintaka}.

In the second step we performed a characterization by PDC  to investigate the phase-matching and spectral properties of the source. The waveguide  was pumped with 200 ns long pulses, which were extracted from an external cavity laser (Toptica DL Pro) by an acoustooptical modulator. After a blocking filter suppressing the pump wave, the photons were launched into a fiber coupled acoustooptical tunable filter with a spectral resolution of $\approx$1.5~nm~\cite{ATOF} and subsequently detected with an InGaAs-single photon detection module (idQuantique 201). The waveguide was temperature stabilized at temperatures around 160~$^o$C (with 0.1~$^o$C stability) to prevent optical damage due to photorefraction.

%
\begin{figure}[b]
\centering\includegraphics[width=12cm]{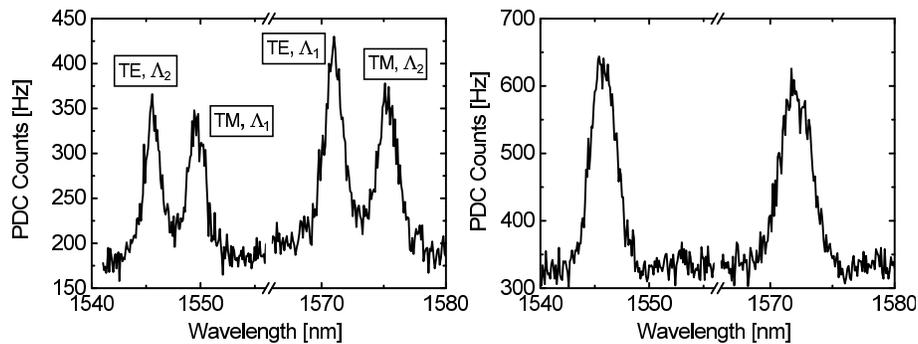}
 \caption{\label{PDC_spectra}Results of spectrally resolved PDC measurements with $\lambda_p=780.0~$nm. Left: PDC spectrum at a sample temperature of $T=159.2~^o$C resulting in separated spectral peaks. Right: Same measurement at $T=156.4~^o$C indicating the correct operation point for polarization entanglement.}
\end{figure}

In \Fig{PDC_spectra} two measured spectra are shown. The measurement yielding the results presented on the left  has been performed at a sample temperature of $T=159.2~^o$C. Four distinct peaks are observable, which correspond from left to right to TE-polarized photons generated in the PDC section with period $\Lambda_2$, TM-polarized from section $\Lambda_1$, TE-polarized from section $\Lambda_1$ and TM-polarized from section $\Lambda_2$. To reach the operation point for polarization entanglement the sample temperature has  been lowered to  $T=156.4~^o$C. At this temperature  the corresponding two peaks overlap completely as illustrated 
in the right diagram of \Fig{PDC_spectra}.

\begin{figure}[htbp]
\centering\includegraphics[width=7cm]{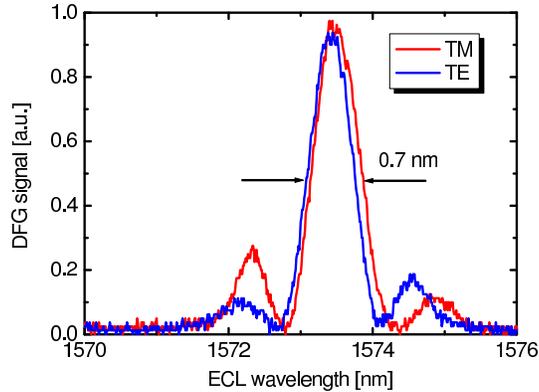}
 \caption{\label{DFG}Measured difference frequency power as function of the wavelength of the ECL acting as signal source for the DFG process. By setting the input polarization either to TE or TM the two different nonlinear process can be probed separately.}
\end{figure}

 The spectral width of the measured curves is about 1.6~nm; but this is essentially determined by the resolution of the acousto-optical filter ($\approx 1.5~$nm). Thus, the spectral bandwidth of the generated PDC signals is not directly measurable with this method. Therefore, we employed difference frequency generation (DFG)  to obtain detailed information on the spectral structure. For this purpose the light from an ECL - tunable around $\lambda=1550$~nm - was superimposed to the pump beam.  The generated DFG signal was recorded as function of the ECL wavelength  as shown in \Fig{DFG}. By setting the input polarization either to TE or TM the conversions belonging to the two different processes could be investigated separately. We found that the shape of the central peak fits quite well and the spectral width (FWHM) is about 0.7 nm. This value is only slightly larger than the theoretically predicted bandwidth of 0.58~nm for a source with 50~mm interaction length. However, we observed more pronounced sidelobes than expected. The deviation from the ideal spectral shape can probably be explained by inhomogeneities of the waveguide along the interaction length. Unfortunately, the shapes of sidelobes of the two curves are different which impact the quality of entanglement as will be discussed in the next section.

%

%
\begin{figure}[tb]
\centering\includegraphics[width=7cm]{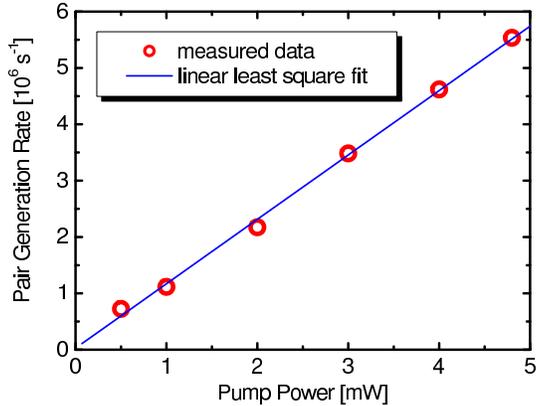}
 \caption{\label{Efficiency}Photon pair generation rate (determined from coincidence measurement) versus coupled pump power. The symbols show the measured data, the solid line is a linear least square fit.}
\end{figure}

%
\begin{figure}[tb]
\centering\includegraphics[width=13cm]{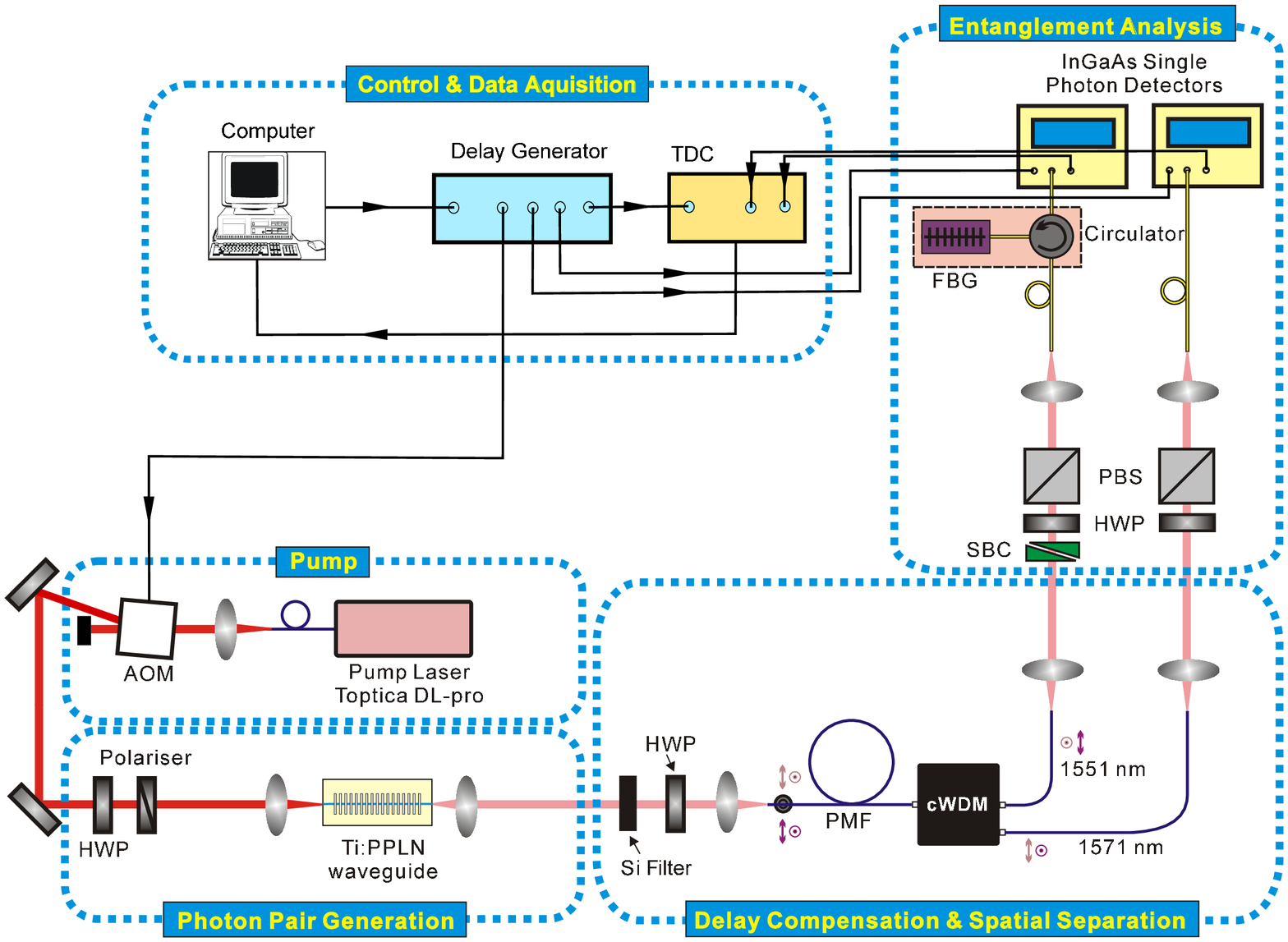}
 \caption{\label{Setup}Experimental setup for the investigation of entanglement. (TDC: time-to-digital converter; AOM: acoustooptical modulator; HWP: half-wave retardation plate; PMF: polarisation-maintaining fiber; cWDM: coarse WDM fiber demultiplexer; SBC: Soleil-Babinet compensator; PBS: polarisation beam splitter; FBG: fiber Bragg grating). }
\end{figure}

In the next step we studied the efficiency of the generation process by using correlation measurements. The generated photons pairs were coupled into a fiber WDM-coupler, spatially separated and detected by single photon detectors. The timing of the registered counting events was then analyzed with a time-to-digital converter. 
We determined the PDC efficiency from the ratio of the coincidence events to the single counts. \Fig{Efficiency} presents the measured pair generation rate in dependence of the pump power. The curve shows a linear relation as predicted. From the slope of this curve a PDC efficiency of $\eta_{PDC}=3 \times 10^{-10}$ is obtained, which is in reasonable good agreement with the predicted efficiency derived from the SH experiments. From this value in combination with the width of the measured spectra a brightness of $B\approx7\times10^3~$pairs/(s$\times$mW$\times$GHz) can be estimated for our source.   

\section{\label{entanglement}Entanglement}
\subsection{Source implementation and characterization}

After the basic characterization of our waveguide structure we set up an experiment for the implementation and characterization of the complete entanglement source. In \Fig{Setup} the overall configuration  is shown. 
The spatial splitting behind the integrated PDC source is accomplished using  
a fiber 1x4 wavelength division demultiplexer (cWDM),  which is a standard  component of telecom fiber networks originally dedicated for coarse WDM applications. We used two output ports centered around 1551 nm and 1571 nm, respectively, with a passband width of 13~nm. The entire cWDM coupler (with specified insertion loss $<$ 2~dB) is equipped with polarization maintaining fibers (PMF). 

The birefringence of the PPLN waveguide leads to a residual temporal distinguishability of the two generation processes resulting in a differential group delay $\Delta\tau$ between TE- and TM-polarized photons. 
To "erase" this distinguishability  a compensating crystal is  usually inserted behind the crystal providing a negative group delay of $-\Delta\tau/2$. (see e.g.\ \cite{Suhara_Review,Suhara1}). In our case we used a PM fiber, which was aligned to couple TE(TM)-polarized photons into the fast (slow) axis mode. The length of the fiber was chosen to provide this compensation and taking into account the additional group delay of the cWDM.

We characterized the entanglement using two rotatable half-wave plates (HWPs) in front of polarization splitters (PBS). A Soleil Babinet compensators (SBC) in one analysis arm enables a fine tuning of the phase $\Phi$. Behind the PBS the photons are coupled into single mode fibers and routed to single photon detectors (idQuantique id201). Their detection events are temporally resolved recorded via a time-to-digital converter enabling the evaluation of  arrival time statistics.

\subsection{Results and discussion}
\begin{figure}[tb]
\centering\includegraphics[width=13.5cm]{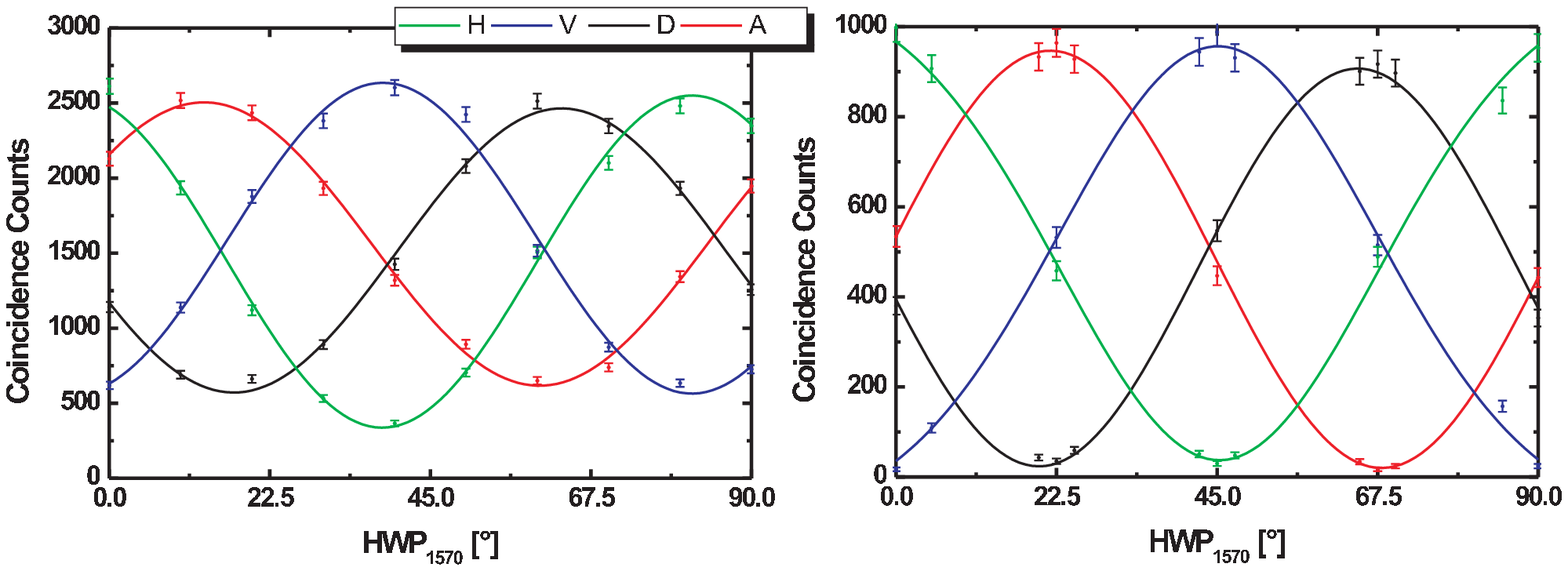}
 \caption{\label{Visibility}Measured visibilities of the entangled photon pair source in the various basis: Left: Visibilities without the additional bandpass filter in the analysis arm and at an input pump power of 16 mW; right: visibilities with additional bandpass filter and input pump power of 6 mW. }
\end{figure}

We performed correlation measurements at various settings of the HWPs. From the measured coincidence counts we obtained net visibilities by substracting accidental coincidences arising from dark counts. In the first experiments we found  initial visibilities in the H/V- and the D/A-bases were only around 70 \% (see \Fig{Visibility},left). 

%

One reason for this relatively low visibility could be attributed to the different spectral shapes of the two processes as revealed by the DFG measurements (\Fig{DFG}). The different spectral distributions (in particular in the sidelobes) lead to a reduced indistinguishability and, thus, caused the reduced visibility. To overcome this problem we inserted an additional bandpass filter consisting of a fiber Bragg grating and a circulator (see \Fig{Setup})  in one of the analysis arms. The FWHM of this filter was about 0.5~nm and, thus, matched exactly the width of the main lobe of the PDC processes. With this filter in place the visibility increased to above 90\%. 

The remaining limitation of the visibility can mostly be explained by multiphoton pair generation. Launching 16 mW pump power into the waveguide results in a highly efficient PDC generation with a high probability to generate multiple pairs. The calculated mean photon number within the 4~ns wide coincidence window is  $\alpha\approx0.08$. Following the model on the impact of multiphoton effects on the visibility described in \cite{Multiphoton} a maximum visibility of about 92.5\% can be expected.  

Reducing the pump power to 6~mW and, thus, the mean photon number in a coincidence window to $\alpha\approx0.03$, increased significantly the visibility. A typical  measurement result is shown in the right diagram of \Fig{Visibility}. A net visibility of more than 95\% has been obtained at the reduced pump power. This result highlights that for all applications of the source a certain trade-off between generation rate and visibility must be selected. 

The final limitation concerning the visibility actually arose from losses within the entire system and a residual polarization dependence of these losses. The measured overall transmisson from the waveguide output facet to the single photon detectors was only about 15~\%.  Most of these losses arose from the incoupling into the fibers, fiber splice losses and the intrinsic losses of the commercial cWDM. Due to all these losses a further reduction of the pump power was not possible to determine the ultimate limit of the visibility because at lower pump powers the count rates dropped below the dark counts preventing a reliable measurement.

An important benchmark of the performance of our entangled source provides the testing of the violation of the Bell inequality. From our measurements we could extract the coincidence data for the various HWP settings required to perform the CHSH test~\cite{CHSH} which enabled us to determine the $S$-parameter to be $S=2.57\pm$0.06. This is a violation of the Bell inequality by more than 9 standards deviations indicating the generation of strongly non-classical photon pairs. 

\section{Conclusions}
We have demonstrated a new source providing polarization-entangled photon-pairs in the telecom regime. Highly efficient photon pair generation is obtained employing PDC in a Ti-diffused PPLN waveguide with an interlaced bi-periodic ferroelectric domain pattern which facilitates type II phase-matching for two wavelength combinations. In combination with standard fiber optical components non-degenerated polarization entangled photon pairs are generated in a fully integrated setup.

A  brightness of the waveguide PDC source as high as $B\approx7\times10^3~$pairs/(s$\times$mW$\times$GHz) could be demonstrated. Our measured visibilities of up to 95\%  were  mainly limited by residual losses in the source and the measurement system. The excellent source characteristics are also verified by  the violation of the Bell inequality
with $S=2.57\pm$0.06.

There is still potential left to improve the device. In particular the free space coupling between the waveguide end-facet and the PMF of the cWDM coupler can be replaced by a direct waveguide-fiber coupling to reduce the losses and to improve the 
compactness of the device. The well matched sizes of Ti-indiffused waveguide modes and fiber modes should allow an efficient coupling  with significantly reduced losses.  Additionally, the cWDM fiber coupler might be replaced by a dense WDM-fiber demultiplexer to narrow the transmission band and, thus, improve the indistinguishabilty of photon pairs generated by the two processes. Finally, a completely fiber pigtailed version seems to be feasible providing high quality entangled photon pairs from a versatile ``Plug\&Play" source.  

\section*{Acknowlegdement}
We thank Raimund Ricken and Viktor Quiring for the waveguide fabrication and Dr. Andreas Christ for many helpful discussions. We acknowledge the funding of parts of this work by the European Space Agency (ESA).

\end{document}